\documentclass{emulateapj}
\usepackage{amstext,amsmath,natbib}
\usepackage{epsfig}

\shorttitle{Interstellar X-ray Spectroscopy II}
\shortauthors{JUETT ET AL.}
\slugcomment{Accepted for Publication in the Astrophysical Journal}

\begin{document}

\newcommand{\XM}{{\em XMM}}
\newcommand{\Ch}{{\em Chandra}}
\newcommand{\eps}{{\rm erg\,s^{-1}}}
\newcommand{\epcs}{{\rm erg\,cm^{-2}\,s^{-1}}}
\newcommand{\cts}{{\rm count\,s^{-1}}}
\newcommand{\jsc}{Paper I}

\title{High-Resolution X-ray Spectroscopy of the Interstellar Medium
  II: Neon and Iron Absorption Edges}

\author{Adrienne~M.~Juett\altaffilmark{1},
Norbert~S.~Schulz\altaffilmark{2},
Deepto~Chakrabarty\altaffilmark{2,3} and
Thomas~W.~Gorczyca\altaffilmark{4}}

\altaffiltext{1}{Department of Astronomy, University of Virginia,
Charlottesville, VA 22903; ajuett@virginia.edu} 
\altaffiltext{2}{Kavli Institute for Astrophysics and Space Research,
Massachusetts Institute of Technology, Cambridge, MA 02139}
\altaffiltext{3}{Department of Physics, Massachusetts Institute of
Technology, Cambridge, MA 02139}
\altaffiltext{4}{Department of Physics, Western Michigan University,
Kalamazoo, MI 49008}

\begin{abstract}
We present high-resolution spectroscopy of the neon $K$-shell and iron
$L$-shell interstellar absorption edges in nine X-ray binaries using
the High Energy Transmission Grating Spectrometer (HETGS) onboard the
{\em Chandra X-ray Observatory}.  We found that the iron absorption is
well fit by an experimental determination of the cross-section for
metallic iron, although with a slight wavelength shift of
$\approx$20~m\AA.  The neon edge region is best fit by a model that
includes the neutral neon edge and three Gaussian absorption lines.
We identify these lines as due to the $1s$-$2p$ transitions from
\ion{Ne}{2}, \ion{Ne}{3}, and \ion{Ne}{9}.  As we found in our oxygen
edge study, the theoretical predictions for neutral and low-ionization
lines all require shifts of $\approx$20~m\AA\/ to match our data.
Combined with our earlier oxygen edge study, we find that a best fit
O/Ne ratio of 5.4$\pm$1.6, consistent with standard interstellar
abundances.  Our best fit Fe/Ne ratio of 0.20$\pm$0.03 is
significantly lower than the interstellar value.  We attribute this
difference to iron depletion into dust grains in the interstellar
medium.  We make the first measurement of the neon ionization fraction
in the ISM.  We find \ion{Ne}{2}/\ion{Ne}{1}$\approx$0.3 and
\ion{Ne}{3}/\ion{Ne}{1}$\approx$0.07.  These values are larger than is
expected given the measured ionization of interstellar helium.  For
\ion{Ne}{9}, our results confirm the detection of the hot ionized
interstellar medium of the Galaxy.
\end{abstract}


\section{Introduction}
High-resolution X-ray spectral observations from the {\em Chandra
X-ray Observatory} and {\em XMM-Newton} offer a new tool with which to
study the interstellar medium (ISM).  The ISM affects X-ray spectra in
two ways: photoelectric absorption, particularly at low energies
(0.1--10~keV), and scattering by dust grains, producing X-ray halos.
At X-ray energies, absorption features are primarily from the
excitation and ionization of inner-shell ($K$-shell) electrons,
although for high-$Z$ elements like iron, $L$-shell absorption edges
are also detectable.  Measurements of the ISM absorption features in
the spectra of X-ray binaries allow us to study the abundances and
ionization fractions for a large number of elements, analogous to
optical and ultraviolet observations of stars that measure ISM
absorption features.  Additionally, high-resolution X-ray spectral
measurements provide a new method for determining the elemental
depletion of the ISM and composition of interstellar dust.

The first attempt to measure ISM absorption edges in the X-rays used
the {\em Einstein} Focal Plane Crystal Spectrometer \citep{sc86} and
found evidence for the \ion{O}{1} $1s$-$2p$ ($K\alpha$) absorption
resonance and a possible \ion{O}{2} edge.  Recent \Ch\/ and \XM\/
results have revealed more complex structure around the oxygen
$K$-shell absorption edge \citep[hereafter
\jsc]{pbv+01,scc+02,tfm+02,ddk+03,jsc04}.  All of the high-resolution
work points to a complex system of absorption features at or near the
oxygen edge, which is not the simple step function used in absorption
models.  Similar high-resolution structure is expected at all of the
absorption edges.

\begin{figure*}
\centerline{\epsfig{file=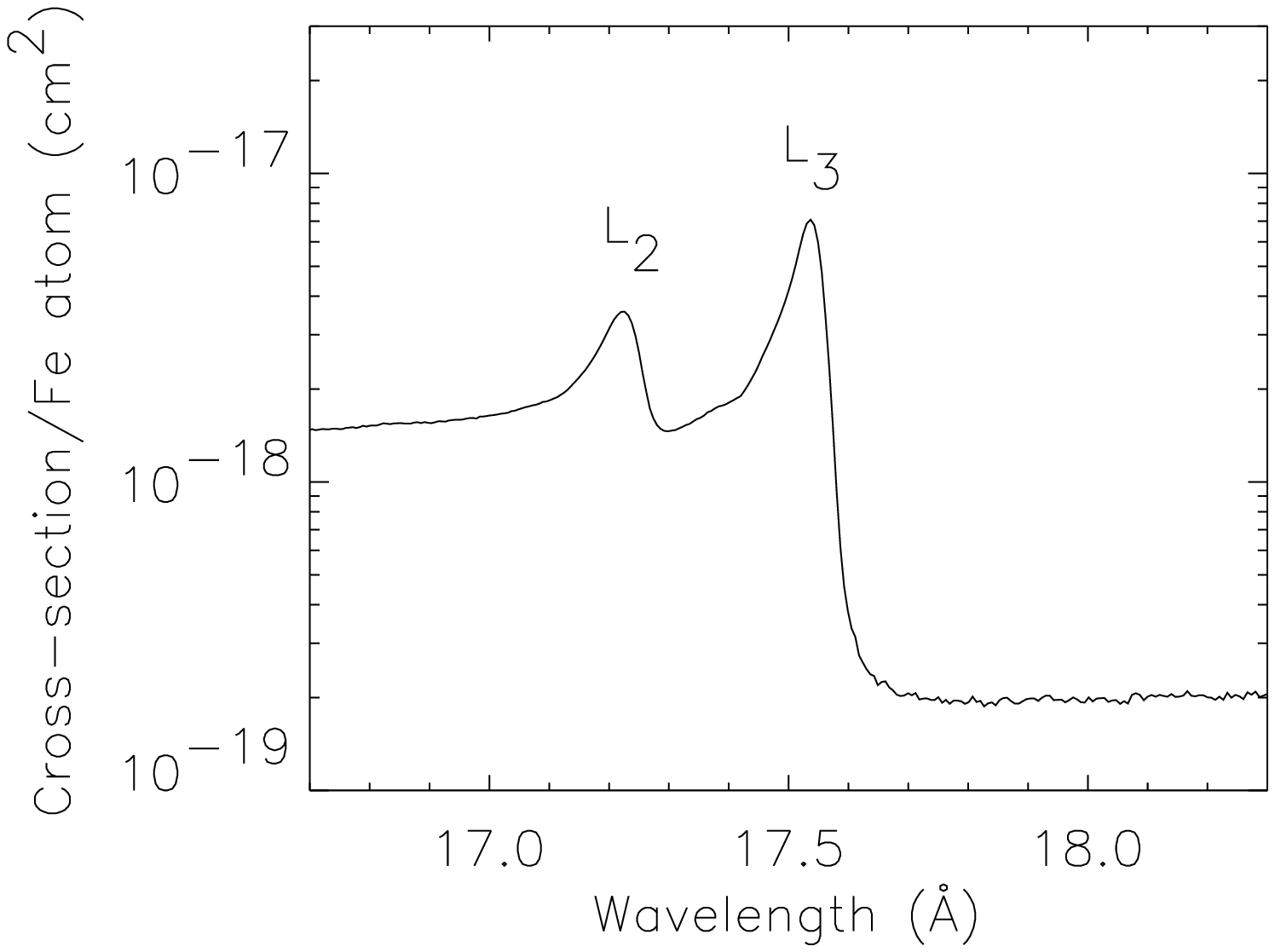, width=0.5\linewidth}\epsfig{file=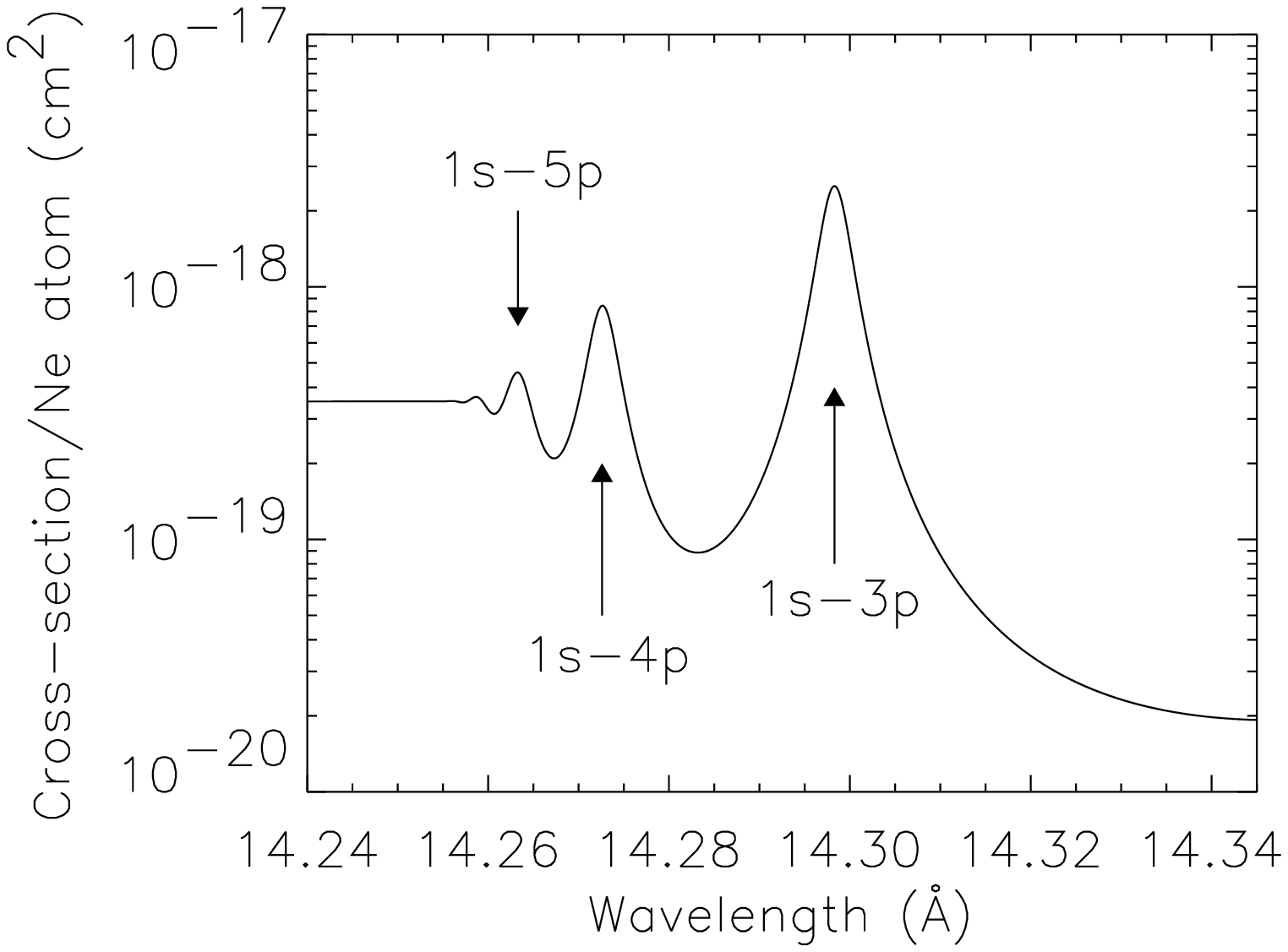, width=0.5\linewidth}}
\caption{{\em Left panel:} Iron $L$-shell cross-section as a function
of wavelength for metallic iron \citep{kk00}.  The positions of the
$L_2$ and $L_3$ edges are marked.  {\em Right panel:} Neon $K$-shell
cross-section as a function of wavelength \citep{g00}.  The positions
of the $1s$-$3p$, $1s$-$4p$, and $1s$-$5p$ transitions are marked.
Only the $1s$-$3p$ transition is resolved by {\em Chandra}.}
\label{fig:cross}
\end{figure*}

Previous results from {\em Einstein} and {\em EXOSAT} grating
observations of the bright low mass X-ray binary (LMXB) Sco X-1 point
to an underabundance of oxygen and an overabundance of nitrogen along
the line of sight to the source \citep{ksc84,bml+85}.  Similarly,
using the abundances of \citet[${\rm O}/{\rm H} = 8.51\times
10^{-4}$]{ag89}, \citet{wop+04} found that the spectrum of the Crab
pulsar was best-fit when oxygen was underabundant.  When the newer
abundances of \citet[${\rm O}/{\rm H} = 4.90\times 10^{-4}$]{wam00}
were used, oxygen was not required to be underabundant, although a
slightly lower than nominal oxygen abundance was preferred, ${\rm
O}/{\rm H} = (3.33\pm0.25) \times 10^{-4}$.  We note that the
measurement of the oxygen abundance relative to hydrogen is misleading
since the X-ray band can not directly measure the hydrogen column
density.  Instead, the measurement is relative to the absorption from
other metals which are converted to an equivalent hydrogen column
density using standard abundances.  \citet{tfm+02} compared the oxygen
and neon column densities measured in the X-ray to the hydrogen column
density from radio observations along the line of sight to Cyg X-2.
Both the neon and oxygen abundances were found to be $\approx3/4$ the
solar abundances of \citet{ag89} but in line with newer estimates of
ISM abundances \citep[e.g.,][]{wam00}.  Finally, \citet{umm+05} found
an overabundance relative to oxygen for magnesium, silicon, sulfur,
and possibly iron \citep[assuming solar abundance ratios from][]{ag89}
in the ISM absorption along the line of sight to three bright neutron
star LMXBs.  While many X-ray binaries have absorption which is well
described by standard abundances, some systems have shown unusual
abundance ratios which have been attributed to material local to these
systems \citep{pbv+01,jpc01,lrr+02,jc03}.  Therefore, an accurate
measurement of the relative abundances along many lines of sight is
necessary in order to interpret data for sources which show unusual
abundance ratios.

We also mention a new result that found a significantly larger Ne/O
ratio than in standard solar or ISM models.  Using the X-ray emission
lines from nearby solar-like stars, \citet{dt05} found an average Ne/O
ratio of 0.41, or 2.7 times greater than the standard ISM or solar
abundances.  If the Sun has a similar Ne/O ratio then this would solve
the solar helioseismology problem \citep[see][]{bbs05}.  We note that
this work has sparked some controversy and claims of smaller Ne/O
ratios in the Sun were quickly made \citep{snr+05,agg+05}.  In either
case, it is not readily apparent that solar coronal measurements
should equal the ISM abundances.

It was recognized early in the history of X-ray spectroscopy that the
composition of dust grains could be probed by spectroscopic
measurements of absorption edges \citep{m70}.  Absorption edges from
molecules include X-ray absorption fine structure (XAFS) due to the
influence of nearby atoms on the absorption process \citep{woo95}.
Such features would distinguish between the molecular and atomic
contributions to the edge structure.  In addition, the resonance
structure of the absorption edges is expected to be shifted, relative
to the atomic position, in molecular absorption edges due to the
molecular binding energy.  The edges from magnesium, silicon, and iron
are expected to be dominated by contributions from molecular, as
opposed to atomic, forms given the high levels of depletion measured
($X_{\rm dust}/X_{\rm total} \gtrsim 95$\%), while carbon, oxygen, and
nitrogen are only mildly depleted \citep[$X_{\rm dust}/X_{\rm total}
\lesssim 60$\%;][]{ss96}.  Several authors have claimed detection of
the $1s$-$2p$ transition in iron oxides \citep{pbv+01,scc+02,tfm+02},
although we showed that this resonance coincides with the $1s$-$2p$
transition from \ion{O}{2} making the identification of iron oxides
debatable (\jsc).  \citet{lrr+02} claimed a tentative detection of
XAFS in the silicon absorption edge of the X-ray binary
GRS~1915$+$105.  Similarly \citet{umm+05} claimed that the silicon and
magnesium edges in the spectra of three LMXBs show XAFS consistent
with the expected structure from silicates.  In addition, depletion of
elements in grains can reduce the effective cross-section for
absorption due to screening \citep[see e.g.,][]{wam00}.  This effect
will lead to deviations in the measured abundances from the true
interstellar values.

Along with neutral and molecular absorption edges, features from
ionized species are also expected.  We showed that low ionization
lines from oxygen are found in the spectra of a number of X-ray
binaries (\jsc).  From these lines, a measurement of the large-scale
ionization fraction of oxygen in the ISM is made.  \citet{mrf+04}
identified \ion{Ne}{2} and \ion{Ne}{3} absorption lines in the spectra
of two black-hole X-ray binaries.  They attributed these to material
local to the systems, although such lines may arise in the ISM similar
to the oxygen lines.  Highly ionized lines may also be expected from
the hot ionized phase of the ISM.  \citet{fmt+04} claim the detection
of an \ion{O}{7} absorption line in the spectrum of the LMXB
4U~1820$-$30, which they attribute to the hot ISM.  Recently,
\citet{yw05} detected \ion{Ne}{9} absorption in the spectra of a
number of LMXBs.  Low-ionization lines can be used to better
understand the relationship between the warm ionized and warm neutral
phases of the ISM.  The high-ionization features in Galactic LMXBs
probe the hot ionized phase of the ISM and have direct implications
for the detection of $z=0$ highly ionized oxygen and neon absorption
features in the spectra of quasars, which have been claimed to be due
to the local group warm-hot intergalactic medium
\citep[e.g.,][]{nze+03} or alternatively a hot Galactic corona
\citep{sws+03}.

In this paper, we study the absorption edges of neon ($K$-shell) and
iron ($L$-shell) in nine bright X-ray binaries.  These edges are most
prominent in systems with Galactic hydrogen column densities of
$10^{21}$--$10^{22}$~cm$^{-2}$.  Many of the X-ray binaries used in
this work were also part of the oxygen edge study, allowing us to
measure the relative abundances of oxygen, iron, and neon.  These
elements also sample a variety of expected depletion levels: neon
should have no depletion, oxygen is mildly depleted, and iron is
heavily depleted.  We will look for signatures of depletion by
comparing the abundance ratios of the elements.  Finally, we will
measure the ionization of neon, using the identified low- and
high-ionization absorption features.

\vspace{0.1in}
\section{Expected Structure of the Iron and Neon Absorption Edges}
In \jsc, we presented an overview of the basic atomic physics
necessary to understand $K$-shell absorption in the X-ray band.  We
defer the reader to the discussion presented there and focus in the
following specifically on the iron $L$-shell and neon $K$-shell
absorption edges.

The iron edge located at 17.5~\AA\/ is due to the excitation and
ionization of $L$-shell ($n=2$) electrons.  The high-resolution
structure of the edge is resolved into three distinct edges, $L_1$,
$L_2$, and $L_3$.  The $L_1$ edge at 14.7~\AA, due to transitions from
the $2s$ level, is the weakest of the edges and we neglect it in our
study since our data are not sensitive enough to measure this
absorption.  The $L_2$ and $L_3$ edges, due to transitions from the
$2p_{1/2}$ and $2p_{3/2}$ levels, are located at 17.2 and 17.5~\AA,
respectively (see Table~\ref{tab:theory} and Figure~\ref{fig:cross}).

To model the cross-section from Fe-$L$, we use the measurement of
metallic iron by \citet{kk00}.  A comparison of the \citet{kk00}
cross-section with lower-resolution determinations
\citep{hgd93,vyb+93,cc95} finds that the values compare to within 15\%
at a wavelength of 16.8~\AA, where the high-resolution structure is no
longer a factor.  Other analyses have found that the \citet{kk00}
measurement reproduces the high-resolution structure of the iron edge
quite well \citep[e.g.,][]{loc+01,scc+02}.

We expect the cross-section from iron in dust to differ in two
important aspects.  First, the shape and position of the edge can
differ depending on the binding of iron within a molecule.  The two
most common forms of iron in molecules are ferrous (Fe$^{2+}$) and
ferric (Fe$^{3+}$).  Spectroscopic studies of the Fe-$L$ edge reveal
that the position of the $L_3$ edge varies from 17.517~\AA\/ for
ferrous iron to 17.475~\AA\/ for ferric iron \citep{vl02}.  Molecules
with a mix of the two forms show a double-peaked profile.  The $L_2$
edge position and structure also varies \citep{vl02}.  Olivines, a
favored form of interstellar dust \citep[see e.g.,][]{d03c}, contains
pure ferrous iron.  Iron oxides can contain ferrous (FeO), ferric
(Fe$_2$O$_3$), or ferrous ferric iron (Fe$_3$O$_4$), depending on the
composition.  Therefore, determination of the position and shape of
the $L_3$ edge can yield information on the composition of
interstellar dust.

The second difference is the strength of the cross-section for
absorption.  In grains, the optical depth can be large ($\sim$1),
therefore absorption will occur primarily on the surface.  Atoms
located in the inner regions of the grains are shielded by the outer
layers.  This produces a net reduction in the cross-section for
elements located in dust grains.  We have estimated this effect for
iron, following the description in \citet{wam00}, and find that the
cross-section for absorption of iron in dust grains is 88\% of the
cross-section for gaseous iron.  We note that this calculation
requires a number of assumptions, including the relative abundances
and depletions of the interstellar elements (see
Appendix~\ref{app:a}).  A more detailed study of these assumptions is
presented in Section~\ref{sec:feabund}.

\begin{deluxetable}{lcc}
\tabletypesize{\footnotesize}
\tablewidth{0pt}
\tablecaption{Iron and Neon Edge and Line Positions}
\tablehead{ & \colhead{Predicted} & \colhead{Measured\tablenotemark{a}} \\ 
  \colhead{Feature} & \colhead{$\lambda$ (\AA)} & \colhead{$\lambda$ (\AA)}
}
\startdata
\multicolumn{3}{c}{Kortright \& Kim 2000} \\ \tableline
Fe-$L_{3}$ edge\tablenotemark{b} & 17.537 & 17.498$\pm$0.003 \\
Fe-$L_{2}$ edge\tablenotemark{b} & 17.226 & 17.188$\pm$0.003 \\ \tableline
\multicolumn{3}{c}{van Aken \& Liebscher 2002} \\ \tableline
Fe$^{2+}$-$L_3$ edge\tablenotemark{b} & 17.517 & 17.498$\pm$0.003 \\
Fe$^{2+}$-$L_2$ edge\tablenotemark{b} & 17.206 & 17.188$\pm$0.003 \\
Fe$^{3+}$-$L_3$ edge\tablenotemark{b} & 17.475 & 17.498$\pm$0.003 \\
Fe$^{3+}$-$L_2$ edge\tablenotemark{b} & 17.156 & 17.188$\pm$0.003 \\ \tableline
\multicolumn{3}{c}{Gorczyca 2000} \\ \tableline
\ion{Ne}{1} $1s$-$3p$\tablenotemark{c} & 14.298 & 14.295$\pm$0.003 \\ 
\ion{Ne}{1} $1s$-$4p$ & 14.273 & \\ 
\ion{Ne}{1} $1s$-$5p$ & 14.263 & \\ \tableline
\multicolumn{3}{c}{Gorczyca \& McLaughlin 2005} \\ \tableline
\ion{Ne}{2} $1s$-$2p$ & 14.605 & 14.608$\pm$0.002 \\
\ion{Ne}{3} $1s$-$2p$ & 14.518 & 14.508$\pm$0.002 \\ \tableline
\multicolumn{3}{c}{Behar \& Netzer 2002} \\ \tableline
\ion{Ne}{2} $1s$-$2p$ & 14.631 & 14.608$\pm$0.002 \\
\ion{Ne}{3} $1s$-$2p$ & 14.526 & 14.508$\pm$0.002 \\
\ion{Ne}{9} $1s$-$2p$ & 13.448 & 13.4439$\pm$0.0013
\enddata
\label{tab:theory}
\tablenotetext{a}{Measured in this work.}
\tablenotetext{b}{Position measured at wavelength of maximum
absorption.}
\tablenotetext{c}{Position of $1s$-$3p$ transition will be coincident
with edge wavelength when using a standard edge model.}
\end{deluxetable}

The neon $K$-shell ($n=1$) absorption edge is located at 14.3~\AA\/
\citep{g00}.  The neon edge, like oxygen, is composed of a series of
resonance lines ending at the series limit or ionization energy.
Neon, and higher $Z$ elements, do not have a $1s$-$2p$ transition due
to a closed $2p$ shell in the neutral atomic ground state
configuration.  \citet{g00} calculated the cross-section for neutral
neon absorption and found that his results closely matched
experimental determinations of the cross-section (see
Figure~\ref{fig:cross}).  The resolution of \Ch\/ and \XM\/ is near
the limit needed to resolve the $1s$-$3p$ resonance line from the rest
of the edge structure of neon, therefore we expect that when using
standard edge models the measured edge positions will be consistent
with the position of the $1s$-$3p$ resonance (see
Table~\ref{tab:theory}).

\begin{deluxetable*}{lrccccccc}
\tabletypesize{\tiny} 
\tablewidth{0pt} 
\tablecaption{Observation Log}
\tablehead{\colhead{Source Name} & \colhead{ObsID} & 
  \colhead{Observation Date} & \colhead{Time (ks)} & \colhead{$l$} & 
  \colhead{$b$} & \colhead{Dist. (kpc)} & \colhead{$\left|z\right|$ (pc)} &
  \colhead{$N_{\rm O}$ ($10^{18}$ cm$^{-2}$)\tablenotemark{a}}
}
\startdata
4U~1820$-$30  & 1021 & 2001-07-21 & 9  & 2.79 & $-$7.91 & 7.6\tablenotemark{b} 
     & 1050 & 1.31$^{+0.20}_{-0.14}$ \\
              & 1022 & 2001-09-12 & 9  & & & & & \\
GX 9$+$9      & 703  & 2000-08-22 & 21 & 8.51 & $+$9.04 & \nodata & \nodata & 
     2.2$^{+1.2}_{-0.6}$ \\
Ser X-1       & 700  & 2000-06-13 & 76 & 36.12 & $+$4.84 & 
     9.5--12.7\tablenotemark{c} & 800--1070 & \nodata \\
Cyg X-1       & 107  & 1999-10-19 & 9  & 71.33 & $+$3.07 & 
     2.0\tablenotemark{d} & 110 & 6.3$\pm1.4$ \\
              & 3407 & 2001-10-28 & 21 & & & & & 2.4$\pm$0.3 \\
              & 3724 & 2002-07-30 & 14 & & & & & 2.9$\pm$0.3 \\
Cyg X-2       & 1102 & 1999-09-23 & 29 & 87.33 & $-$11.32 & 
     11.4--15.3\tablenotemark{c} & 2240--3000 & 1.1$\pm$0.2 \\
4U~1636$-$53  & 105  & 1999-10-20 & 29 & 332.91 & $-$4.82 & 
     3.7--4.9\tablenotemark{c} & 310--410 & 2.6$^{+1.0}_{-0.3}$ \\ 
              & 1939 & 2001-03-28 & 26 & & & & & \\
GX 339$-$4    & 4420 & 2003-03-17 & 74 & 338.94 & $-$4.33 & 
     $>$6\tablenotemark{c} & $>$450 & \nodata \\
4U~1735$-$44  & 704  & 2000-06-09 & 24 & 346.05 & $-$6.99 & 
     8.0--10.8\tablenotemark{c} & 970--1310 & 3.4$\pm$1.2 \\
GX 349$+$2    & 715  & 2000-03-27 & 9  & 349.10 & $+$2.75 & \nodata & \nodata 
     & \nodata \\
              & 3354 & 2002-04-09 & 26 & & & & &
\enddata 
\label{tab:obs}
\tablenotetext{a}{Oxygen column density measurement from \citet{jsc04}.}
\tablenotetext{b}{\citet[and references therein]{kdi+03}.}
\tablenotetext{c}{\citet[and references therein]{jn04}.}
\tablenotetext{d}{\citet{gzp+99}.}
\end{deluxetable*}

We also expect ionized neon to be present in the ISM.  The primary
transition we will find is from the $1s$-$2p$ transition in ionized
neon.  Unlike neutral neon, the $1s$-$2p$ transition is available for
ionized forms.  \citet{bn02} calculated the wavelengths and oscillator
strengths of the strongest $1s$-$np$ (primarily $1s$-$2p$) transitions
for a variety of elements and ionizations.  The authors estimate that
the errors on the calculated values are $\lesssim$0.2\% for
wavelengths and $\lesssim$10\% for oscillator strengths and
autoionization rates.  They also note that the accuracy increases for
higher ionization states.  We have also calculated photoelectric
absorption cross-sections for \ion{Ne}{2} and \ion{Ne}{3} using the
same method as for neutral neon \citep{gm05}.  We list the predicted
positions of the \ion{Ne}{2}, \ion{Ne}{3}, and \ion{Ne}{9} lines from
both sources in Table~\ref{tab:theory}.  Comparing the predicted
wavelengths for the \ion{Ne}{2} $1s$-$2p$ transition, we find that the
two calculations differ by $\approx$25~m\AA, on the order of the error
given by \citet{bn02}.  The oscillator strengths are consistent within
a few percent.  We note that the published natural width for
\ion{Ne}{2} in \citet{bn02} is incorrect and that the correct value is
$2.710$$\times$$10^{14}$~s$^{-1}$ (Behar 2005, priv. comm.).

Using the full cross-sections for the neon species, we have checked
that the neutral neon edge does not overlap with transitions from
\ion{Ne}{2} and \ion{Ne}{3}.  The \ion{Ne}{2} resonance series limit
is at $\approx$13.7~\AA\/ while the \ion{Ne}{3} limit is at
$\approx$13.3~\AA.  No resonance lines from \ion{Ne}{2} or \ion{Ne}{3}
are coincident with the neutral neon edge.  The edges are easily
distinguishable given the spectral resolution of {\em Chandra}.

\section{Observations and Spectral Fitting Procedure}
Our study was based on archival observations of nine bright X-ray
binaries obtained with {\em Chandra} using the High Energy
Transmission Grating Spectrometer (HETGS) in combination with the
Advanced CCD Imaging Spectrometer \citep[ACIS;][]{cdd+05}.  The HETGS
spectra are imaged by ACIS, an array of six CCD detectors.  The
HETGS/ACIS combination provides both an undispersed (zeroth order)
image and dispersed spectra from the gratings.  The various orders are
sorted using the intrinsic energy resolution of the ACIS CCDs.  The
HETGS carries two types of transmission gratings: the Medium Energy
Gratings (MEGs) with a range of 2.5--31~\AA\/ (0.4--5.0~keV) and the
High Energy Gratings (HEGs) with a range of 1.2--15~\AA\/
(0.8--10.0~keV).  The first-order MEG (HEG) spectrum has a spectral
resolution of $\Delta\lambda=$ 0.023~\AA\/ (0.012~\AA).  The absolute
wavelength accuracy of the MEG (HEG) is $\pm$0.011~\AA\/ (0.006~\AA).

In Table~\ref{tab:obs}, we list the observations used in this
analysis.  The data sets were reduced using the standard CIAO
threads\footnote{http://asc.harvard.edu/ciao/threads/}.  In some
cases, the zeroth-order data was not telemetered.  For these
observations, we estimated the zeroth-order position by finding the
intersection of the grating arms and readout streak.  After spectral
extraction, the accuracy of the estimated zeroth-order position was
verified by comparing the wavelengths of strong instrumental edges in
both plus and minus sides of the spectra.  Events collected during
thermonuclear X-ray bursts were excluded from this analysis.  Since
the individual observations of 4U~1820$-$30 and 4U~1636$-$53 were
brief, the spectra from multiple observations were combined for each
source in order to improve statistics.  Detector response files (ARFs
and RMFs) were created for the $+$1 and $-$1 MEG and HEG spectra.  The
ARFs include the standard correction for the time-dependent change in
the response due to a contaminant on ACIS.  We then added the $+$1 and
$-$1 orders to produce a combined MEG and HEG spectra from each
source.  Background spectra were also extracted.  The MEG data were
used for both the iron and neon edge fits, while the HEG data were
used for the neon edge when reasonable signal-to-noise was found.
All fitting was performed in count space.  In this paper, we present
the flux-corrected spectra in order that the reader may better
visualize the interstellar features.  These flux spectra are created
in ISIS using the tool {\tt flux\_corr} which yields model-independent
flux spectra.

\subsection{Iron Edge Fit}\label{sec:fefit}
\begin{figure*}
\centerline{\epsfig{file=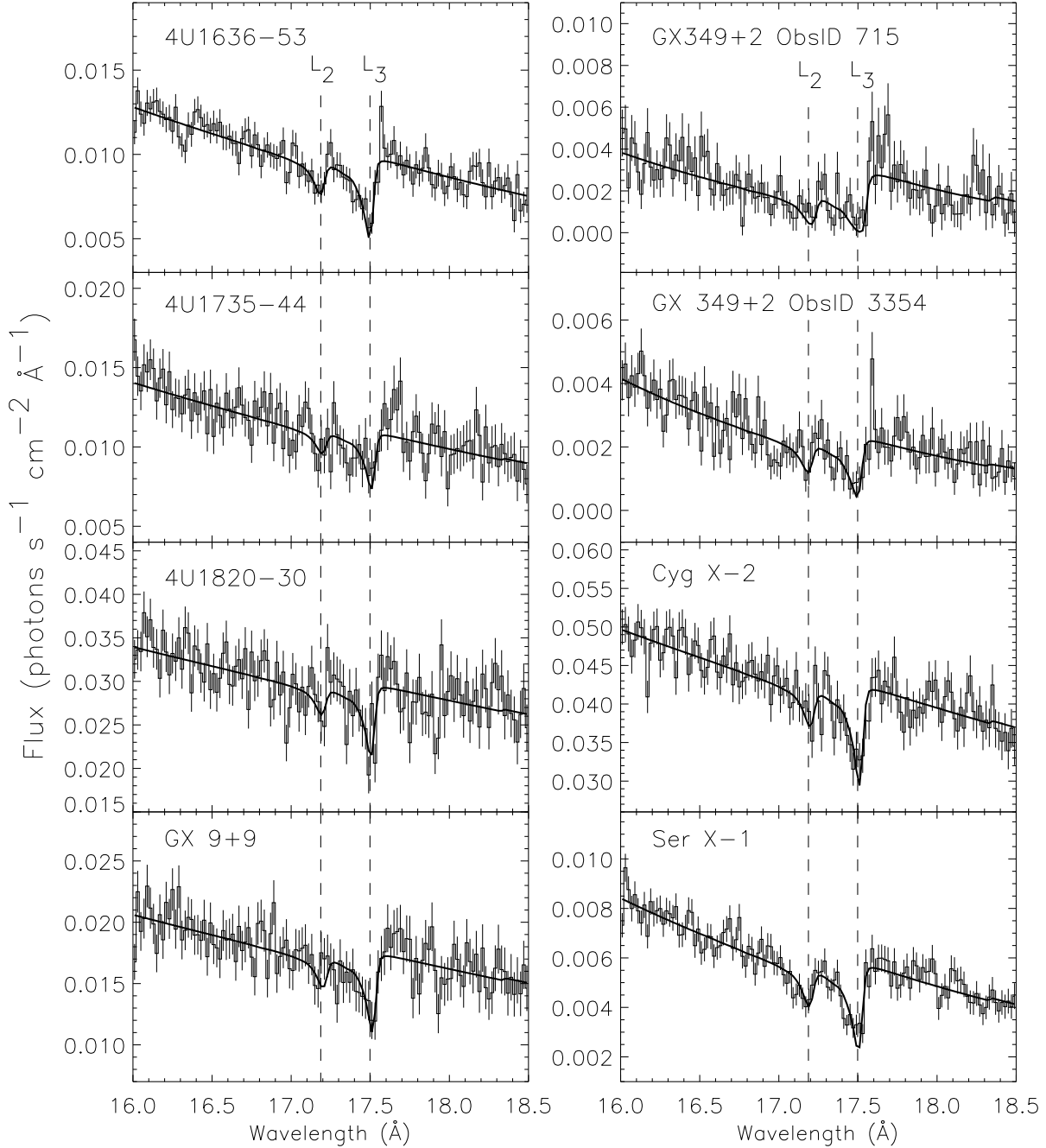,width=0.9\linewidth}}
\caption{Flux spectra of the iron $L$-shell absorption region for the
seven neutron star LMXBs included in this study.  The data have been
binned by a factor of two for illustrative purposes.  The dashed lines
indicate the positions of the $L_2$ and $L_3$ edges.}
\label{fig:fens}
\end{figure*}

\begin{figure}
\centerline{\epsfig{file=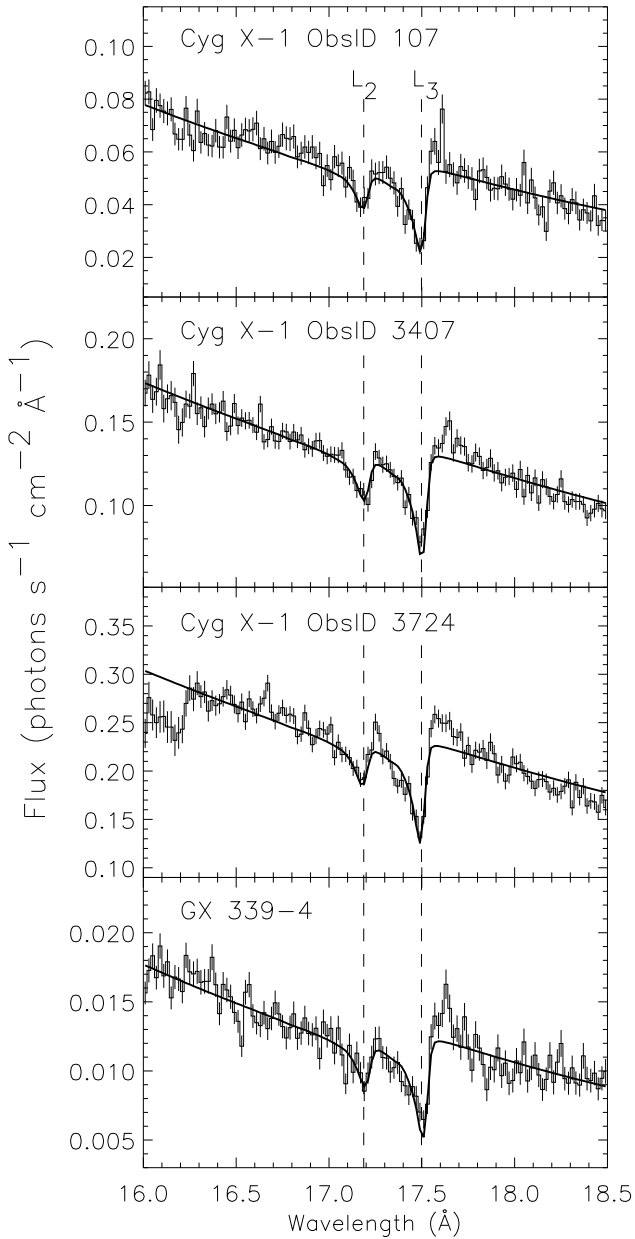,width=\linewidth}}
\caption{Flux spectra of the iron $L$-shell absorption region for the
two black hole X-ray binaries included in this study.  The data have
been binned by a factor of two for illustrative purposes.  The dashed
lines indicate the positions of the $L_2$ and $L_3$ edges.}
\label{fig:febh}
\end{figure}

We fit only the 16--18.5~\AA\/ wavelength range and binned the data to
ensure at least 10 counts per bin.  An absorbed power-law continuum
model was used for all sources.  We allowed the power-law model
parameters to vary during the fits.  The {\tt tbvarabs} absorption
model \citep{wam00} was used with the iron abundance set to zero to
allow for an explicit modeling of the iron absorption.  The equivalent
hydrogen column density ($N_{\rm H}$) of the {\tt tbvarabs} model was
set to equal a multiplicative factor times the iron column density
($N_{\rm Fe}$), where the multiplicative factor was the inverse of the
iron ISM abundance of \citet{wam00}.  We modeled the iron absorption
edge by using a custom multiplicative model based on the iron
cross-section of \citet{kk00}.  The model parameters were $N_{\rm Fe}$
and the edge wavelength, defined as the position of the maximum
cross-section of $L_3$ edge.  Both were allowed to vary in the fits.
The best-fit $N_{\rm Fe}$ and edge wavelengths are given in
Table~\ref{tab:fe}.  All error estimates are 90\%-confidence levels,
unless otherwise noted, and do not include instrumental errors.
Figures~\ref{fig:fens} and \ref{fig:febh} show the data and best-fit
model for each spectrum.

\begin{deluxetable}{lcc}
\tabletypesize{\footnotesize} 
\tablewidth{0pt} 
\tablecaption{Fe Edge Parameters}
\tablehead{\colhead{Source} & \colhead{$\lambda$ (\AA)} & 
  \colhead{$N_{\rm Fe}$ (10$^{16}$ cm$^{-2}$)}}
\startdata
4U~1820$-$30          &   17.505$\pm$0.017 &    5.1$\pm$1.9 \\
GX 9$+$9              &   17.512$\pm$0.015 &      7$\pm$2 \\
Ser X-1               &   17.502$\pm$0.010 &     15$\pm$3 \\
Cyg X-1 ObsID 107     &   17.495$\pm$0.009 &   14.7$^{+2.5}_{-1.9}$ \\
Cyg X-1 ObsID 3407    &   17.501$\pm$0.007 &   10.6$^{+0.7}_{-1.1}$ \\   
Cyg X-1 ObsID 3724    &   17.491$\pm$0.004 &    9.6$\pm$0.9 \\
Cyg X-2               &   17.510$\pm$0.012 &    5.6$\pm$1.4 \\
4U~1636$-$53          &   17.495$\pm$0.010 &     10$\pm$2 \\
GX 339$-$4            &   17.505$\pm$0.007 &   14.4$^{+2.2}_{-1.8}$ \\
4U 1735$-$44          &    17.51$\pm$0.02 &       6$\pm$3 \\
GX 349$+$2 ObsID 715  &    17.52$\pm$0.04 &   69$^{+11}_{-39}$ \\
GX 349$+$2 ObsID 3354 &    17.49$\pm$0.02 &   26$^{+17}_{-9}$     
\enddata
\label{tab:fe}
\end{deluxetable}

\subsection{Neon Edge Fit}
\begin{figure*}
\centerline{\epsfig{file=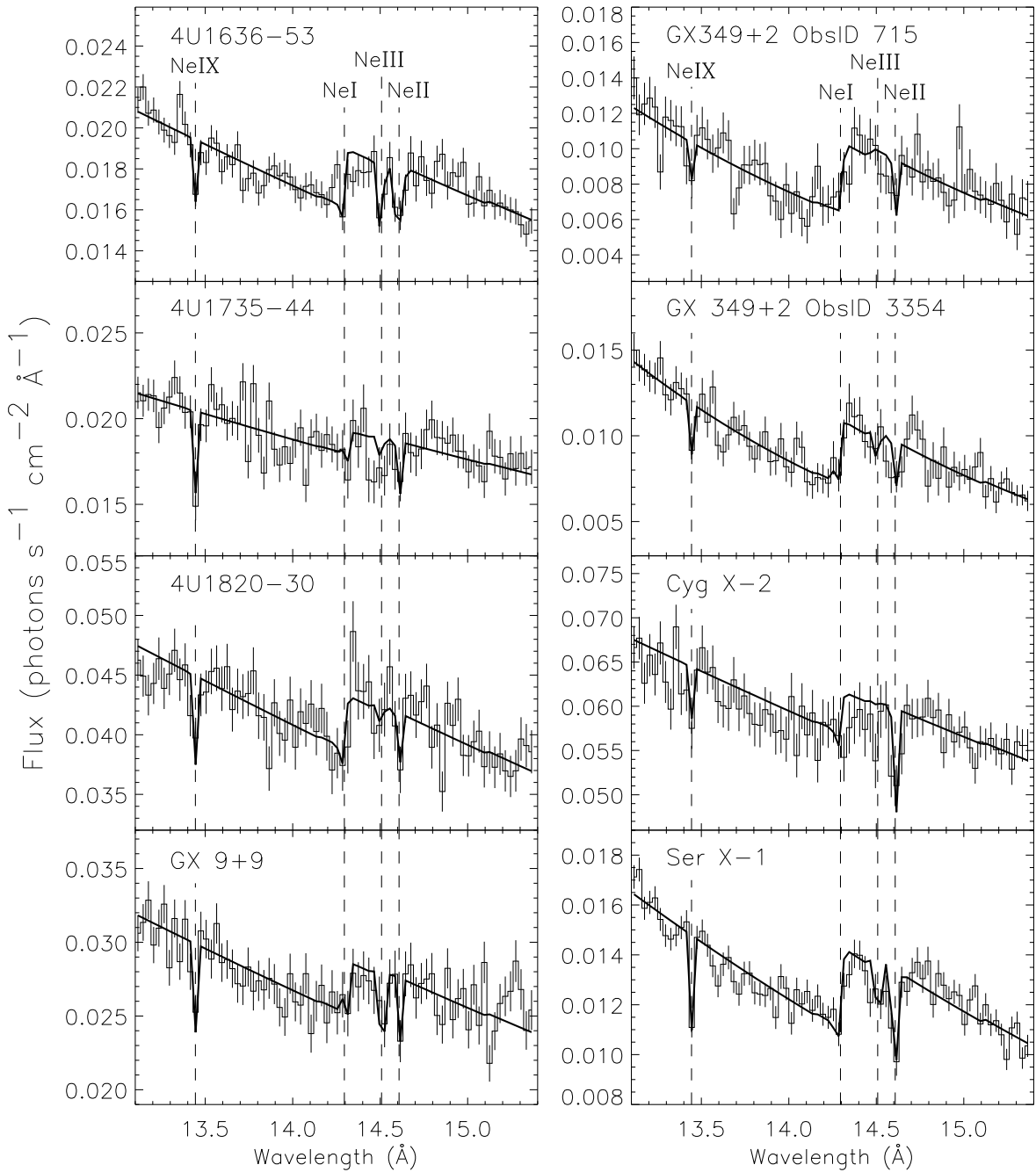,width=0.9\linewidth}}
\caption{Flux spectra of the neon $K$-shell absorption region for the
seven neutron star LMXBs included in this study.  The data have been
binned by a factor of three for illustrative purposes.  The dashed
lines indicate the positions of the identified features.}
\label{fig:nens}
\end{figure*}

\begin{figure}
\centerline{\epsfig{file=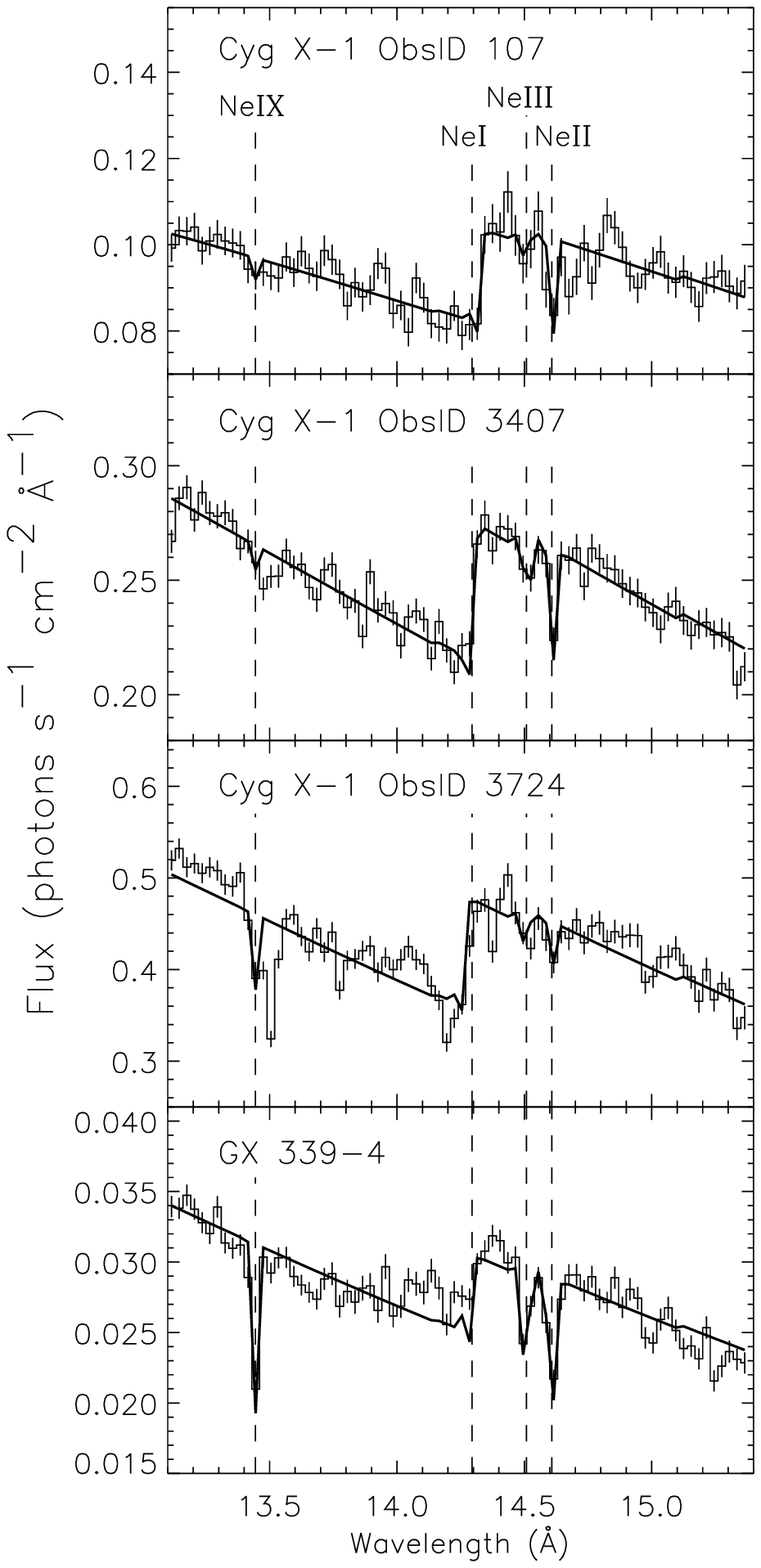,width=\linewidth}}
\caption{Flux spectra of the neon $K$-shell absorption region for the
two black hole X-ray binaries included in this study.  The data have
been binned by a factor of three for illustrative purposes.  The
dashed lines indicate the positions of the identified features.}
\label{fig:nebh}
\end{figure}

We fit only the 13.1--15.4~\AA\/ wavelength range and again binned the
data to ensure at least 10 counts per bin.  The continuum modeling
used the same procedure as for iron (see \S~\ref{sec:fefit}).  We
modeled the neutral neon absorption edge by using a custom
multiplicative model based on the neon cross-section of \citet{g00}.
The model parameters were $N_{\rm Ne}$ and the edge wavelength,
defined as the position of the $1s$-$3p$ resonance.  Both were allowed
to vary in the fits.  Also included were three Gaussian absorption
lines to measure the absorption by \ion{Ne}{2}, \ion{Ne}{3}, and
\ion{Ne}{9}.  Initial fits fixed the position of these lines to those
found for GX~339$-$4 \citep{mrf+04} and the widths (sigma) to
0.005~\AA\/ (the instrument resolution).  The earlier GX~339$-$4
result represents the best position identification for the
low-ionization lines and was therefore used as the reference.  Our
results confirm the position of these lines.  For sources where all
three absorption lines were significantly detected, as opposed to only
upper limits on their flux, we then refit the data, allowing the line
positions and widths to vary.  The best-fit parameters were used to
determine the equivalent widths (EWs) of the lines.  The best-fit
$N_{\rm Ne}$ and edge wavelengths are given in Table~\ref{tab:needge},
while the best-fit line positions, widths, and EWs are given in
Table~\ref{tab:neline}. Figures~\ref{fig:nens} and \ref{fig:nebh} show
the data and best-fit model for each spectrum.

\vspace{0.1in}
\section{Results and Analysis}

We now compare the results of our analysis for the sample of sources.
From this we will determine the line and edge positions, the relative
abundances of the neutral and ionized species, and depletion
fractions.

\begin{deluxetable}{lcc}
\tabletypesize{\footnotesize} 
\tablewidth{0pt} 
\tablecaption{Ne Edge Parameters}
\tablehead{\colhead{Source} & \colhead{$\lambda$ (\AA)} & 
  \colhead{$N_{\rm Ne}$ (10$^{17}$ cm$^{-2}$)}}
\startdata
4U~1820$-$30          & 14.294$\pm$0.012 &   3.3$\pm$1.2 \\
GX 9$+$9              & 14.304$^{+0.008}_{-0.013}$ & 3.9$\pm$1.3 \\
Ser X-1               & 14.293$\pm$0.007 &          7.67$^{+0.19}_{-0.7}$ \\
Cyg X-1 ObsID 107     & 14.317$\pm$0.007 &   7.1$\pm$1.0 \\
Cyg X-1 ObsID 3407    & 14.293$\pm$0.004 &   7.4$^{+0.7}_{-0.3}$ \\
Cyg X-1 ObsID 3724    & 14.256$\pm$0.012 &   8.6$\pm$0.7 \\
Cyg X-2               & 14.294$\pm$0.010 &   2.3$^{+0.9}_{-0.3}$ \\
4U~1636$-$53          & 14.291$\pm$0.012 &   4.9$\pm$1.1 \\
GX 339$-$4            & 14.281$\pm$0.012 &   6.08$\pm$0.13 \\
4U 1735$-$44          & 14.315$\pm$0.013 &   2.2$^{+1.3}_{-0.8}$ \\
GX 349$+$2 ObsID 715  & 14.30$^{+0.05}_{-0.03}$ &  14.40$^{+0.11}_{-0.8}$ \\
GX 349$+$2 ObsID 3354 & 14.284$^{+0.030}_{-0.013}$ &  13$\pm$3 
\enddata
\label{tab:needge}
\end{deluxetable}

\begin{deluxetable}{lccc}
\tabletypesize{\tiny} 
\tablewidth{0pt} 
\tablecaption{Ne line parameters}
\tablehead{\colhead{Source} & \colhead{$\lambda$ (\AA)} & 
  \colhead{$\sigma$ (\AA)} & \colhead{EW (m\AA)}}
\startdata
4U~1820$-$30 &   14.5065 (fixed)  &  0.005 (fixed) &  $>~ -$5 \\
             &   14.6067 (fixed)  &  0.005 (fixed) &  $>~ -$7 \\
             &   13.4420 (fixed)  &  0.005 (fixed) &  $-4.8^{+1.9}_{-2.9}$ \\
GX 9$+$9     &   14.511$\pm$0.013 &  0.013$\pm$0.010 & $-9^{+4}_{-7}$ \\
             &   14.610$\pm$0.010 &  $<$0.009      & $-$5$\pm$3 \\
             &   13.438$\pm$0.008 &  $<$0.010      & $-6^{+2}_{-3}$ \\
Ser X-1      &   14.510$\pm$0.004 &  $<$0.014      & $-7.2^{2.3}_{-1.7}$ \\
             &   14.606$\pm$0.006 &  0.012$\pm$0.006 & $-$12$\pm$3 \\
             &   13.446$\pm$0.002 &  $<$0.007      & $-$7.7$\pm$1.4 \\
Cyg X-1      &   14.5065 (fixed)  &   0.005 (fixed) & $>~ -$6 \\
ObsID 107    &   14.6067 (fixed)  &   0.005 (fixed) & $-$7$\pm$3 \\
             &   13.4420 (fixed)  &   0.005 (fixed) & $>~ -$3.6 \\
Cyg X-1      &   14.511$\pm$0.005 &   $<$0.013     & $-$4$\pm$2 \\
ObsID 3407   &   14.612$\pm$0.004 &   0.009$\pm$0.006 & $-$6$\pm$3 \\
             &   13.4420 (fixed)  &   0.005 (fixed) & $>~ -$2 \\
Cyg X-1      &   14.5065 (fixed)  &   0.005 (fixed) & $-$3.0$\pm$1.9 \\
ObsID 3724   &   14.6067 (fixed)  &   0.005 (fixed) & $-$3.2$\pm$1.9 \\
             &   13.4420 (fixed)  &   0.005 (fixed) & $-$5.3$\pm$1.8 \\
Cyg X-2      &   14.5065 (fixed)  &  0.005 (fixed) & $>~ -3.5$ \\
             &   14.6067 (fixed)  &  0.005 (fixed) & $-$6.2$\pm$2.7 \\
             &   13.4420 (fixed)  &  0.005 (fixed) & $>~ -$5 \\
4U~1636$-$53 &   14.502$\pm$0.008 &  0.013$\pm$0.010 & $-$7$\pm$4 \\
             &   14.601$\pm$0.012 &  $<$0.04   &  $-$10$\pm$7 \\
             &   13.442$\pm$0.006 &  $<$0.015 & $-$5$\pm$2 \\
GX 339$-$4   &   14.506$\pm$0.003 &  0.008$\pm$0.003 & $-$9.4$\pm$1.3 \\
             &   14.607$\pm$0.003 &  0.011$\pm$0.003 & $-$12$\pm$2 \\
             &   13.4425$\pm$0.0019 & 0.005$\pm$0.002 & $-$11.6$\pm$1.3 \\
4U 1735$-$44 &   14.5065 (fixed)  &  0.005 (fixed) &  $>~ -$6 \\
             &   14.6067 (fixed)  &  0.005 (fixed) &  $>~ -$12 \\
             &   13.4420 (fixed)  &  0.005 (fixed) & $-7^{+2}_{-4}$ \\
GX 349$+$2   &   14.5065 (fixed)  &  0.005 (fixed) & $>~ -$13 \\
ObsID 715    &   14.6067 (fixed)  &  0.005 (fixed) & $>~ -$37 \\
             &   13.4420 (fixed)  &  0.005 (fixed) & $>~ -$20 \\
GX 349$+$2   &   14.5065 (fixed)  &  0.005 (fixed) & $>~ -$21 \\
ObsID 3354   &   14.6067 (fixed)  &  0.005 (fixed) & $>~ -$27 \\
             &   13.4420 (fixed)  &  0.005 (fixed) & $>~ -$18 
\enddata
\label{tab:neline}
\end{deluxetable}

\begin{figure}
\centerline{\epsfig{file=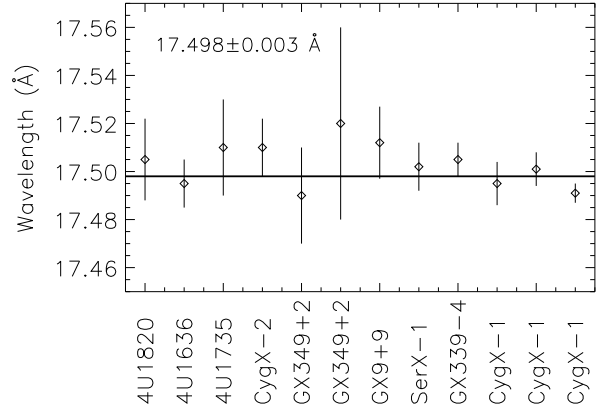,width=\linewidth}}
\caption{Best fit position of the iron $L_3$ absorption edge, defined
as the position of maximum cross-section.  The weighted mean value of
the line position is given.  All quoted errors are statistical only
and do not include the instrumental wavelength error of
$\pm$0.011~\AA\/ (FWHM).}
\label{fig:feedge}
\end{figure}

Our data are well fit by the metallic iron cross-section of
\citet{kk00}, although requiring a small shift in the position of the
$L_3$ edge.  This is comparable to previous studies of the iron edge
\citep[e.g.,][]{loc+01,scc+02}.  The weighted mean position of the
Fe-$L_3$ edge (taken as the position of maximum optical depth) is
17.498$\pm$0.003~\AA\/ (errors are 90\% confidence and do not account
for the absolute instrumental wavelength accuracy of $\pm$0.011~\AA;
see Figure~\ref{fig:feedge}).  Even including the absolute error, this
value is significantly different from the measured positions of both
the ferrous and ferric edges and lies between the two values
\citep{vl02}.  This may suggest that the ISM contains a combination of
ferrous and ferric forms of iron.  But we caution that the energy
resolution of the laboratory data \citep[0.8~eV$=$20~m\AA\/ at
17.5~\AA;][]{vl02} is large enough to make either value consistent
with our measurement.  The spacing of the $L_2$ and $L_3$ edges is
well fit by our model and is more consistent with ferrous iron.  This
would imply that the laboratory measurement of the wavelength of
ferrous iron is off by $\approx$19~m\AA.

If both ferrous and ferric forms are present in the ISM, we would
expect a double-peaked structure in the combined edge.  We searched
the residuals from the best-fit for our data.  A negative residual at
17.40~\AA\/ was found, although with low significance, $<3\sigma$.
Even if true, the wavelength of the feature is too small to be
attributable to a contribution from ferric iron, which we would expect
at 17.45~\AA\/ if our best-fit position is from ferrous iron.  In
addition, a check of the effective area versus wavelength for the
sources with the most prominent residuals found sharp variations in
the effective area, a signature of hot pixels or columns, in the iron
wavelength range.  We also note the positive residuals just high of
the edge position (17.6~\AA).  We know of no atomic or molecular
features that would produce such a residual.  It is also possible that
these residuals are due to hot pixels or columns.

\begin{figure}
\centerline{\epsfig{file=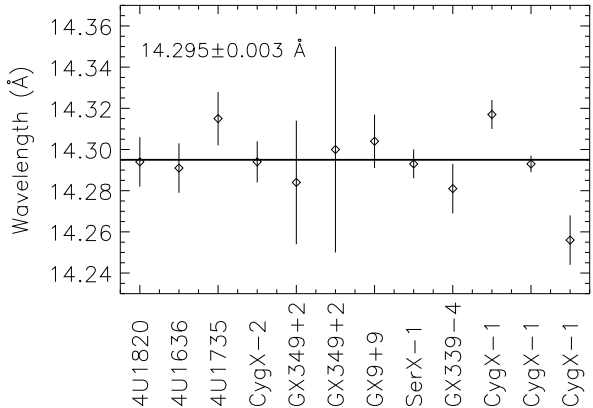,width=\linewidth}}
\caption{Best fit position of the $1s$-$3p$ transition from
\ion{Ne}{1}.  The weighted mean value of the line position is given.
All quoted errors are statistical only and do not include the
instrumental wavelength error of $\pm$0.011~\AA\/ (FWHM).}
\label{fig:needge}
\end{figure}

For the neutral neon edge, we find a weighted mean position of the
$1s$-$3p$ transition of 14.295$\pm$0.003~\AA\/ (see
Figure~\ref{fig:needge}).  This is consistent with laboratory
measurements \citep[e.g.,][]{cac+99}.  We note that the theoretical
position of \citet{g00} was calibrated to this laboratory measurement.
For the ionized neon lines, we find a weighted mean position of
14.608$\pm$0.002~\AA\/ for \ion{Ne}{2}, 14.508$\pm$0.002~\AA\/ for
\ion{Ne}{3}, and 13.4439$\pm$0.0013~\AA\/ for \ion{Ne}{9}.  The
\ion{Ne}{9} position is consistent with theoretical and observational
measurements, accounting for the absolute instrument error.  The
\ion{Ne}{2} and \ion{Ne}{3} lines, however, are $\approx$20~m\AA\/ off
of the theoretical predictions.  This is within the quoted errors for
the theoretical calculations \citep{bn02}.  In this case, the error is
more significant for the theory than the observations.  We encountered
a similar effect when identifying the neutral and low-ionization lines
from oxygen (\jsc).  We note that our best-fit positions for the
\ion{Ne}{2} and \ion{Ne}{3} lines are consistent with the results of
\citet{mrf+04}.

\subsection{Relative Abundances of Oxygen, Iron, and Neon}\label{sec:feabund}
Using the column density measurements, including the oxygen column
density measurements of \jsc, we can now directly measure the
abundance ratios in the X-ray domain.  This measurement is a function
not only of the interstellar abundances, but also of the depletion of
the elements into dust grains.  We have chosen to use neon as the
calibration point for these measurements, thereby providing O/Ne and
Fe/Ne ratios, for two reasons.  First, as a noble gas, neon should not
be depleted.  Second, as the lowest wavelength edge of the set
presented here, it lies near the middle of the {\em Chandra}
wavelength band.  Observations of higher column density sources will
show edges from neon, magnesium, and silicon.  Therefore, abundance
ratios relative to neon will allow us to directly compare the ISM
abundances of various elements, e.g., oxygen and silicon, which could
not be done with a single source in the X-ray waveband.

We calculated the abundance ratios (by number) for the entire sample
and for the neutron star (NS) LMXBs only.  While the black hole
systems in our sample are known to show intrinsic features
\citep{scc+02,ftz03,mws+03,mrf+04}, the NS LMXBs show no emission or
absorption features that are intrinsic to the system.  NS LMXBs which
do show intrinsic features are not included in this study.  We
therefore take the NS subsample to be indicative of the ISM properties
and show the black hole systems relative to this subsample.  We note
that the inclusion of the black hole systems in this study may seem
unnecessary given their intrinsic features, but they provide the
highest signal-to-noise spectra allowing us to make more accurate
measurements of the line and edge positions.

\begin{figure}
\centerline{\epsfig{file=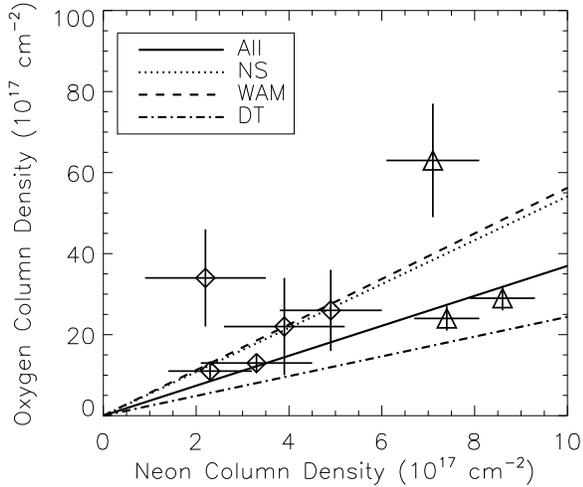,width=\linewidth}}
\caption{Comparison of the neutral oxygen and neon column densities
for our sample of sources.  The best fit O/Ne ratio is plotted for all
sources (solid line) and the neutron star LMXBs only (dotted line).
We have overplotted the ISM O/Ne ratio of \citet[dashed line]{wam00}
and the stellar coronal measurement of \citet[dashed-dotted
line]{dt05}.  The best fit value for the NS LMXBs in is good agreement
with the ISM value.  None of our values are consistent with the
\citet{dt05} O/Ne ratio.}
\label{fig:nevso}
\end{figure}

In Figure~\ref{fig:nevso}, we compare the neon and oxygen column
densities for our sample.  For all the systems, we find a O/Ne ratio
of 3.7$\pm$0.3.  When we restrict our sample to the NS systems, we
find O/Ne$=$5.4$\pm$1.6.  The difference in the measurements is driven
by the low oxygen column measurements for two of the Cyg X-1 datasets.
We note that while the oxygen column for Cyg X-1 shows a factor of 2
difference between the three datasets, the neon column density
measurements are all consistent within errors.  This is likely due to
material local to the source (see \jsc\/ and references therein).

Our abundance ratio measurement for the NS systems is consistent with
O/Ne$=$5.6 obtained from the ISM abundances of \citet{wam00}.  This is
also consistent with standard solar abundance ratios
\citep[O/Ne$=$6.6; e.g.,][]{l03,ags05}.  Interestingly, our results
exclude the abundance ratio of \citet{dt05}, O/Ne$=$2.4.  One possible
explanation is that our two studies cover a very different extent in
the Galaxy.  All of our sources are $\gtrsim$2~kpc from the Sun, while
the \citet{dt05} sample of stars are all within 100~pc.

Our best-fit O/Ne ratio also allows for roughly 45\% of the oxygen in
the ISM to be located in grains (see derivation below and
Appendix~\ref{app:a}), although we note that the large error on the
O/Ne ratio allows no real constraint.  More precise measurements of
the O/Ne ratio are needed to probe the depletion in the ISM.

\begin{figure}
\centerline{\epsfig{file=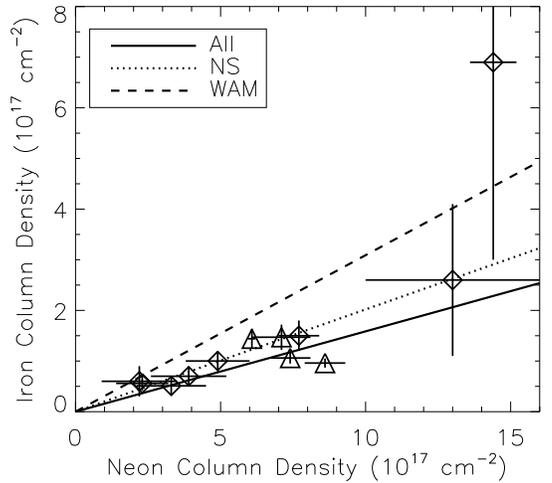,width=\linewidth}}
\caption{Comparison of the neutral iron and neon column densities for
our sample of sources.  The best fit Fe/Ne ratio is plotted for all
sources (solid line) and the neutron star LMXBs only (dotted line).
We have overplotted the ISM Fe/Ne ratio of \citet[dashed line]{wam00}.
Our best fit Fe/Ne ratio is significantly lower than the ISM value.
We attribute this to depletion of iron into dust grains in the ISM.}
\label{fig:nevsfe}
\end{figure}

The Fe/Ne ratio, however, shows a large deviation from the
\citet{wam00} ISM abundance ratio of 0.309 (see
Figure~\ref{fig:nevsfe}).  Including all systems, we find
Fe/Ne$=$0.159$\pm$0.013, while for NS systems only,
Fe/Ne$=$0.20$\pm$0.03.  Both measurements are significantly different
than the ISM value.  The sense of the discrepancy is that we are
finding less iron in the ISM than we would expect.  The discrepancy
between the measured and predicted Fe/Ne abundance ratio is likely the
result of depletion in the ISM.  As mentioned previously, atoms in
dust grains have a lower effective cross-section than the same atoms
in gas.  Therefore, using the gas cross-section (as we have) would
cause the column density measurements to be lower than the true value.

Attributing the abundance ratio difference to depletion, we can
estimate the amount of iron located in dust grains.  The total optical
depth of iron, $\tau_{\rm Fe}$, is the sum of the contributions from
gas and dust grains.  The gaseous iron optical depth is the gas phase
cross-section times the column density of iron in the gas phase, given
by $\sigma_{\rm gas} N_{\rm Fe,gas}$.  The optical depth from iron in
dust is the grain cross-section times the column density of iron in
grains given by $\sigma_{\rm grain} N_{\rm Fe, grain}$.  The depletion
factor, $\beta$, measures the amount of the total iron column density
found in grains, $N_{\rm Fe,grain} = \beta N_{\rm Fe}$.  Similarly,
the gas column density is $N_{\rm Fe,gas} = (1 - \beta) N_{\rm Fe}$.
Finally, the dust cross-section can be related to the gas
cross-section by a factor, $f$, as shown in Appendix~\ref{app:a}.
Combining this we find
\begin{equation}
\tau_{\rm Fe} = \sigma_{\rm gas} (1 - \beta) N_{\rm Fe} + f
\sigma_{\rm gas} \beta N_{\rm Fe} = \sigma_{\rm gas} N_{\rm Fe} (1 -
\beta + f\beta)~.
\end{equation}
Our analysis assumes that $\tau_{\rm Fe}=\sigma_{\rm gas} N_{\rm Fe}$.
Therefore we can relate our measured iron column density, $N_{\rm
Fe,meas}$, to the true column density by $N_{\rm Fe,meas} = N_{\rm Fe}
(1 - \beta + f\beta)$.  Since we do not know the true $N_{\rm Fe}$, we
must use abundances ratios to measure depletion.  Using abundance
relative to neon allows us to avoid additional depletion effects that
would be present with oxygen.  The relationship between the measured
Fe/Ne abundance ratio and the ISM ratio is given by
\begin{equation}\label{eqn:dep}
({\rm Fe/Ne})_{\rm meas} = (1 - \beta + f\beta) \times ({\rm
Fe/Ne})_{\rm ISM}~,
\end{equation}
assuming $\beta=0$ for neon.  Given our measured Fe/Ne ratio, the ISM
abundance ratio from \citet{wam00}, and the value of $f$ calculated in
Appendix~\ref{app:a}, we find that $\beta = 2.9\pm0.8$, which is
unphysical since $\beta$ is constrained to the range 0--1.

Obviously, one or more of our assumptions is invalid.  It is possible
that the ISM abundances of either iron or neon is incorrect.
\citet{wam00} claim errors for the ISM abundances of 0.1~dex or
higher.  A neon abundance 0.1~dex larger is enough to account for the
discrepancy and constrain $\beta > 0.5$.  A slightly greater neon
abundance is not at present problematic for our measured O/Ne ratio,
given the large error bars.  Similarly, a lower iron abundance could
also account for the difference.

\begin{figure}
\centerline{\epsfig{file=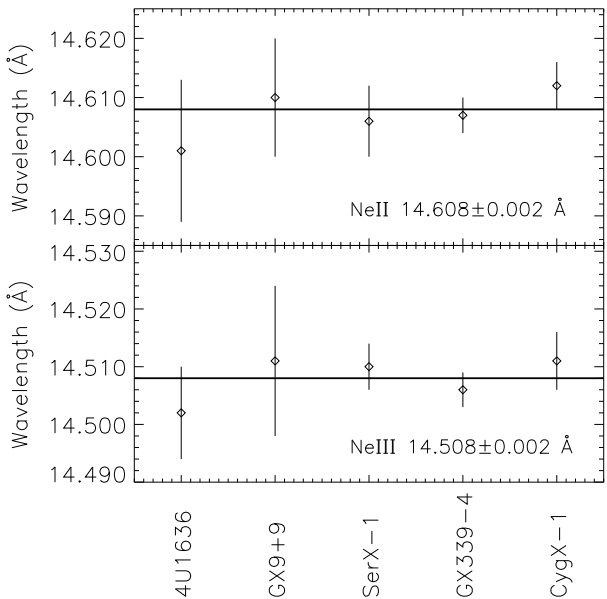,width=\linewidth}}
\caption{Best fit positions of the $1s$-$2p$ transitions from
\ion{Ne}{2} ({\em top panel}) and \ion{Ne}{3} ({\em bottom
panel}). All quoted errors are statistical only and do not include the
instrumental wavelength error of $\pm$0.011~\AA\/ (FWHM).}
\label{fig:lowionne}
\end{figure}

We have also tested the possibility that the dust grain cross-section
could be misestimated.  Assuming the \citet{wam00} abundances, for
various depletion factors, $\beta$$=$0.7--0.95, we find that
Equation~\ref{eqn:dep} gives values of $f$$=$0.4--0.7, significantly
lower than our calculated value.  We have calculated $f$, as shown in
Appendix~\ref{app:a}, for various assumptions, including the dust
model \citep{mrn77,wd01}, abundances, variations in depletion factors
of the astrophysically abundant elements, grain mass density, and the
gas cross-section for iron.  Most of these parameters do not make a
large difference to the value of $f$.

Only the cross-section and grain mass density are able to make a
significant effect, although they must be more than 5 times larger to
bring the measured and predicted results into agreement.  A difference
this large in the iron cross-section is unlikely.  First, we are
already using the maximum value of the cross-section, not the average
value over the edge which would give a lower cross-section.  In
addition, the cross-section of \citet{kk00} compares well with
lower-resolution measurements at wavelengths away from the
high-resolution structure.  Therefore the cross-section difference
between the $L_3$ maximum and the lower wavelength absorption would
have to be substantially greater than found by \citet{kk00}.  Since
our data fit both the high-resolution structure and the lower
wavelength absorption, we would expect to see deviations in our data
from the chosen model if this was the case.  Such deviations are not
found and we therefore conclude that the iron edge model is accurate
in its structure and scale to at least 10\%.

Finally, we address a final assumption in the calculation of $f$,
which is a homogeneous distribution of the elements in grains.
\citet{sf93} suggested that interstellar silicate grains may have a
iron-rich core with a magnesium-mantle.  Such a scenario would provide
more shielding to the iron atoms, producing a lower value of $f$.

\subsection{Ionized Neon Abundances}
The diffuse ISM has four major phases: cold neutral, warm neutral,
warm ionized, and hot ionized \citep{hk87}.  The strength of the
ionized neon lines can be used to constrain the properties of these
phases.  The \ion{Ne}{2} and \ion{Ne}{3} lines are most likely
associated with the warm ionized medium (WIM), while the neutral edges
will be primarily from the warm neutral medium (WNM).  From an
ultraviolet observation of the Galactic halo star HD~93521,
\citet{sf93} found that the WIM coincided with the WNM.  Other studies
\citep{hs99,rtk+95} have found similar results.  This allows us to
make a direct comparison of the column densities from the neutral and
low ionization neon lines from which the ionization of the ISM can be
measured.

\begin{figure}
\centerline{\epsfig{file=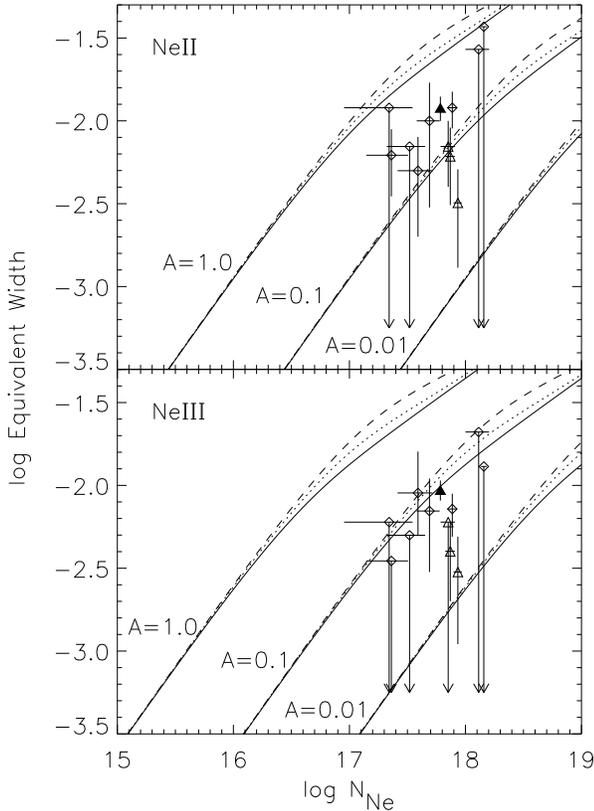,width=\linewidth}}
\caption{Curve of growth for the \ion{Ne}{2} and \ion{Ne}{3}
absorption features.  Plotted are the measured values of the neutral
neon column density, $N_{\rm Ne}$, and the EW in the ionized neon
absorption lines.  The diamonds mark the NS LMXB systems, the open
triangle marks the black hole binary Cyg X-1, and the filled triangle
marks the black hole LMXB GX 339$-$4.  Overlayed are the theoretical
predictions for the relationship between the column density and EW in
the lines.  The solid, dotted, and dashed lines indicate the predicted
EWs for velocity dispersion of 20, 100, and 200~km~s$^{-1}$,
respectively.  Also included are predicted EWs for various abundances
($A = N_{\rm X}/N_{\rm NeI}$) relative to neutral neon.  From these
figures, we can estimate the relative abundances for \ion{Ne}{2} and
\ion{Ne}{3} relative to \ion{Ne}{1}.}
\label{fig:cog}
\end{figure}

The ionized abundances can be estimated from a curve of growth
analysis (see~\jsc).  We used the theoretical cross-section of
Gorczyca (2005; in prep.) to calculate the curve of growth for
\ion{Ne}{2} and \ion{Ne}{3} assuming thermal velocities of 20, 100,
and 200~km~s$^{-1}$.  Using the oscillator strengths and natural
widths of \citet{bn02} gives consistent results.  In
Figure~\ref{fig:cog}, we plot the EW in the ionized lines versus the
neutral neon column density for each source.  Given an ionized
abundance, $A = N_{\rm X}/N_{\rm NeI}$, one can then overlay the
curves of growth calculated from the theoretical line properties.  The
best-fit abundance is determined by comparing the theoretical curves
at different values of A to the data points.  We note that there is a
degeneracy between the thermal velocity and the ionized abundance.
Given the low significance of our EW measurements, we only estimate
the ionized abundances.  We find that
\ion{Ne}{2}/\ion{Ne}{1}$\approx$0.3 and
\ion{Ne}{3}/\ion{Ne}{1}$\approx$0.07.

Our analysis attributes the ionized neon lines to interstellar
material.  In our oxygen edge study, we argued that an interstellar
interpretation is correct, since the EWs of the lines correlated
better with the neutral column density than with the source luminosity
(see \jsc).  The same trend is suggested by the ionized neon lines,
although at much less significance.  Obviously, more accurate
measurements are required to verify this trend.

It is interesting to note that the EWs of the low-ionization neon
lines from GX~339$-$4 give a result in line with the NS LMXB trend.
This suggests that the low and neutral absorption along the line of
sight to GX 339$-$4 is consistent with an interstellar origin.  On the
other hand, the EWs for the low-ionization neon lines from Cyg X-1 do
not follow the same trend.  In the oxygen study, Cyg X-1 showed large
variations in the column densities of the neutral and ionized lines
(\jsc).  We attributed those changes to circumstellar material in the
binary.  The variations in the neon features are less substantial,
which may imply something interesting about the source of the ionizing
flux.

\begin{figure}
\centerline{\epsfig{file=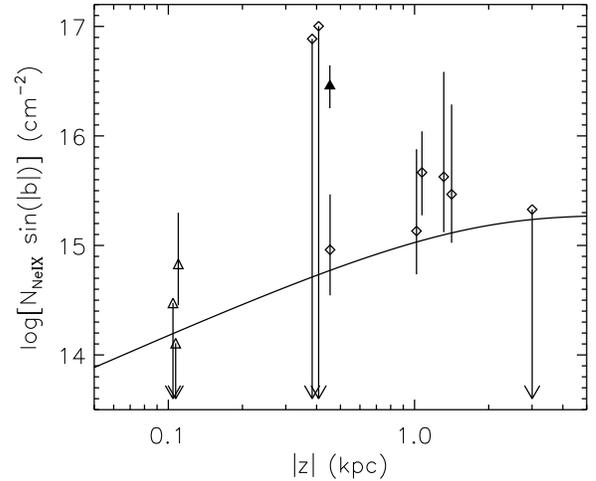,width=\linewidth}}
\caption{\ion{Ne}{9} column density as a function of source distance
above the plane.  Column densities were calculated from the EWs
assuming a gas temperature of $2.4\times10^{6}$~K.  For systems with
no known distance, we assume a distance of 8.5~kpc, while for
GX~339$-$4 we used the lower limit on the distance of 6~kpc.  Slight
shifts in $|z|$ were applied for illustrative purposes.  Overplotted
is the best fit disk distribution of \citet{yw05}.  The diamonds mark
the NS LMXB systems, the open triangle marks the black hole binary Cyg
X-1, and the filled triangle marks the black hole LMXB GX 339$-$4.
Notice that the data for GX~339$-$4 is the most discrepant from the
model, confirming a local contribution to the \ion{Ne}{9} line.}
\label{fig:ne9}
\end{figure}

The \ion{Ne}{9} absorption should originate in the hot phase of the
ISM.  With a higher temperature and lower density, the hot phase will
have a larger scale height than the warm phases which contain the
neutral and low ionization lines.  Therefore, a direct comparison is
not appropriate.  Instead, we follow \citet{yw05} and compare the
\ion{Ne}{9} column densities as a function of $z$, the distance from
the Galactic plane (see Figure~\ref{fig:ne9}).  We overplot their
best-fit disk distribution.  The \ion{Ne}{9} column densities were
calculated from a curve of growth assuming a gas temperature of
$2.4\times10^{6}$~K \citep{yw05}.  Our results compare reasonably well
with their best-fit model, which is not surprising given the large
overlap of sources.  Again, we see variations in the Cyg~X-1 (open
triangles) column densities.  The data point that deviates the most
from the model is from GX~339$-$4 (filled triangle).  This confirms
that a sizable contribution to the \ion{Ne}{9} line is due to a local
contribution, which is reasonable given the detection of other
hydrogenic and helium-like lines in the spectrum of this source
\citep{mrf+04}.

\vspace{0.1in}
\section{Discussion}
We have presented a global study of the neon $K$-shell and iron
$L$-shell absorption in the ISM.  Combined with our previous study of
oxygen absorption (\jsc), we have determined the neutral abundances
ratios between oxygen, iron, and neon.  We find O/Ne$=$5.4$\pm$1.6, in
line with current ISM abundances.  Our Fe/Ne abundance of
0.20$\pm$0.03 is lower than the expected ISM value of 0.309,
suggesting that iron is depleted in the ISM.  We note that unless
depletion is taken into account in X-ray absorption models, a subsolar
iron abundance may be found.  This subsolar iron abundance can be due
only to the ISM and does not {\em a priori} signify an astrophysically
interesting result related to the source properties.

We also detected absorption lines from \ion{Ne}{2}, \ion{Ne}{3}, and
\ion{Ne}{9} in a number of the sources.  We attribute these lines to
the ISM.  A curve of growth analysis found that the large-scale
ionization of neon in the ISM is \ion{Ne}{2}/\ion{Ne}{1}$\approx$0.3
and \ion{Ne}{3}/\ion{Ne}{1}$\approx$0.07.  The implied column
densities in the \ion{Ne}{9} lines are consistent with the model of
\citet{yw05} for all sources except the black hole LMXB GX~339$-$4.

\subsection{Ionization of the Warm ISM}
The ISM of the Galaxy is typically modeled as a multiphase medium
\citep[see e.g.,][]{mo77}.  The phases have different temperatures,
densities, and ionizations.  The following discussion focuses on two
of the phases, the warm neutral and warm ionized mediums (WNM and WIM,
respectively).  These two phases have similar temperatures
($T\sim10^{4}$K) but different hydrogen ionization fractions.  In the
WNM, the ionized hydrogen fraction \ion{H}{2}/H$_{\rm total}\sim 0.1$,
while in the WIM, \ion{H}{2}/H$_{\rm total}\gtrsim 0.8$
\citep[see][and references therein]{hss03}.

Our implied ionization fractions for \ion{Ne}{2} and \ion{Ne}{3} are
greater than we found for oxygen, \ion{O}{2}/\ion{O}{1}$\approx$0.1
and \ion{O}{3}/\ion{O}{1}$\approx$0.05 (\jsc).  With higher first
ionization potentials, we would expect the ionization of neon to track
that of helium in the ISM.  Measurements of the \ion{He}{1}
$\lambda$5876/H$\alpha$ line intensity ratio suggest a helium
ionization fraction $n($\ion{He}{2}$)/n($He$_{\rm total}) \lesssim
0.27$ in regions where the hydrogen is primarily ionized
\citep[i.e. in the WIM;][]{rt95}.  This has been taken to imply that
the helium is primarily neutral in the WIM, since the alternative
explanation, that helium is primarily fully ionized, appears
inconsistent with the low ionization of other elements, particularly
sulfur and nitrogen.

Our measured neon ionization ratios include the contributions from all
phases of the ISM.  In the following, we attribute the neutral and
low-ionization features to the WNM and WIM only.  These phases should
be the dominant contributors.  We would like to estimate the
\ion{Ne}{2}/Ne$_{\rm tot}$ ratio in the WIM to compare with the
helium result.  To do this, we assume that the \ion{Ne}{2} and
\ion{Ne}{3} lines are produced in the WIM only, while the neutral neon
absorption includes contributions from both the WNM and WIM.  Given
the higher ionization potential of neon compared to hydrogen, we
believe this is valid based on the hydrogen ionization in each phase.
We must make two other assumptions: first, the \ion{Ne}{1},
\ion{Ne}{2}, and \ion{Ne}{3} are the dominant contributors to the
total interstellar neon budget, and second, that a significant
fraction ($\gtrsim75$\%) of the neutral neon is found in the WNM, as
opposed to the WIM.  Combining these assumptions, we find
\begin{multline}
\biggl(\frac{{\rm Ne~II}}{{\rm Ne}_{\rm tot}}\biggr)_{\rm WIM} = \\
\frac{({\rm Ne~II/Ne~I})_{\rm tot}}{{\rm Ne~I_{\rm WIM}/Ne~I_{\rm
tot} + (Ne~II/Ne~I)_{\rm tot} + (Ne~III/Ne~I)_{\rm tot}}}~.
\end{multline}
This work directly measures (\ion{Ne}{2}/\ion{Ne}{1})$_{\rm tot}$
and (\ion{Ne}{3}/\ion{Ne}{1})$_{\rm tot}$, and we are assuming
\ion{Ne}{1}$_{\rm WIM}$/\ion{Ne}{1}$_{\rm tot} \lesssim 0.25$.

Our results then yield \ion{Ne}{2}/Ne$_{\rm tot} \gtrsim 0.4$ in the
WIM.  This is different from the helium results, although we note that
the exact values of the ionization fraction of \ion{Ne}{2} and
\ion{Ne}{3} are dependent on the assumed thermal velocity of the ISM,
which may make the difference between the measured ionizations of
helium and neon less significant.  In addition, our measurement of the
\ion{Ne}{3}/\ion{Ne}{1} ratio seems incompatible with standard OB star
models for the ionization of the WIM \citep[see e.g.,][]{shr+00}.
Extra heating has been studied to explain the properties of optical
emission and absorption lines \citep[e.g.,][]{smh00,ed05}, but these
results concentrate on the elements of interest in the optical.
Therefore, neon ionization fractions are not given.  One possible
explanation for our neon ionization results is that the lines of sight
probed by the X-ray studies contain a larger WIM fraction relative to
the WNM.

X-ray studies probe different lines of sight than traditional optical
and ultraviolet work.  Many of our sight lines are towards the center
of the Galaxy and the Galactic bulge.  Optical and ultraviolet studies
concentrate on the Perseus arm of the Galaxy which is roughly
anti-Galactic center ($120^{\circ}<l<150^{\circ}$).  While X-ray
allows us to probe larger distances and cover a greater extent of the
Galaxy, this also makes it difficult to reconcile X-ray results with
other ISM studies.  Future work will use multiwavelength studies of
X-ray binaries to compare X-ray ISM results with optical and
ultraviolet studies.

\subsection{The Warm Absorber in GX 339$-$4}
\citet{mrf+04} identified the \ion{Ne}{2}, \ion{Ne}{3}, and
\ion{Ne}{9} lines in the spectrum of GX~339$-$4 and attributed them to
an AGN-like warm absorber.  Our results show that the \ion{Ne}{9} line
shows a clear excess column density over the expected interstellar
component.  The measured \ion{Ne}{9} column density for GX~339$-$4 is
$\approx$$3.9\times10^{17}$~cm$^{-2}$.  For Galactic latitude of
GX~339$-$4 and a distance of 6~kpc, the ISM model predicts a column
density of $7.8\times10^{15}$~cm$^{-2}$.  While the exact distance to
GX~339$-$4 is unknown, the model column density as a function of
Galactic scale height reaches a maximum value of
$2.5\times10^{16}$~cm$^{-2}$.  From these values, we estimate that
only $\approx$2--6\% of the measured \ion{Ne}{9} column density to
GX~339$-$4 is due to the hot ISM.  This seems reasonable given the
detection of other highly ionized lines in the spectrum of the source.

The properties of the \ion{Ne}{2} and \ion{Ne}{3} lines however, are
completely consistent with the other LMXBs in our sample.  The
measured line widths are all $\approx$20~m\AA\/ off from the
theoretical calculations of \citet{bn02}, within the quoted errors for
the calculations.  Similar wavelength shifts were also required in our
study of the oxygen edge (\jsc).  The line widths are also consistent
with the other sources in our sample, and are more likely indicative
of the resolution of the instrument rather than an astrophysical
origin.  Finally, the implied ionized abundances of \ion{Ne}{2} and
\ion{Ne}{3} from the GX~339$-$4 data are consistent with the NS LMXBs.
If these lines had a local component, we should see deviations from
these relationships.  We note that variations in the EW of these lines
were claimed by \citet{mrf+04b} but we caution that comparisons
between {\em Chandra}/HETG and {\em XMM}/RGS spectra are suspect for
narrow absorption features due to the factor of two difference in the
spectral resolution between the instruments.  Narrow features easily
detectable by {\em Chandra} can be substantially washed out by the RGS
resolution.  We therefore claim that the low-ionization neon features
in GX~339$-$4 are completely consistent with an interstellar origin.

\vspace{-0.2in}\acknowledgements{We thank E. Behar for providing us
with his data and for useful discussions on the relevant atomic
physics.  Support for this work was provided by the National
Aeronautics and Space Administration through the Smithsonian
Astrophysical Observatory contract SV3-73016 to MIT for Support of the
Chandra X-Ray Center, which is operated by the Smithsonian
Astrophysical Observatory for and on behalf of the National
Aeronautics Space Administration under contract NAS8-03060.  TWG was
supported in part by NASA APRA grant NNG0-4GB58G and by NASA SHP SR\&T
grant NNG05GD41G.}

\vspace{0.1in}
\appendix
\section{Derivation of Cross-section for Dust Grains}\label{app:a}

The derivation of the cross-section for dust grains presented in
\citet{wam00} includes the contribution of all elements.  For our
data, we need the iron dust cross-section only.  Therefore, we present
a rederivation of the calculation of the dust cross-section for a
single element, $Z$.

We begin by considering Equation A7 of \citet{wam00}:
\begin{equation} \label{eqn:1}
\tau_{\rm grains} = N_{\rm H} \xi_{\rm g} \int^{\infty}_{0}
\frac{dn_{\rm gr}(a)}{da} \sigma_{\rm geom} [1 - \exp(-\langle \sigma
\rangle \langle N \rangle)] da~,
\end{equation}
where $N_{\rm H}$ is the hydrogen column density, $\xi_{\rm g}$ is the
number of grains per hydrogen atom, $dn_{\rm gr}(a)/da$ is the grain
size distribution assuming spherical grains of size $a$, $\sigma_{\rm
geom} = \pi a^2$ for spherical grains, $\langle \sigma \rangle$ is the
average photoelectric absorption cross-section of the grain material
and $\langle N \rangle$ is the column density in a grain.

For our purposes, we want $\tau_{\rm grains}$ of element $Z$.  $N_{\rm
H}$ can be directly related to the element column density $N_Z$ by
$N_{\rm H} = N_{Z}/A_{Z}$, where $A_{Z}$ is the ISM abundance of
element $Z$.  Since we are considering a single element, $\langle
\sigma \rangle$ is just the gas phase cross-section of that element,
$\sigma_{\rm gas}$.

To calculate $\langle N \rangle$ for a single element, we follow
\citet{wam00}, and assume that the grains have a homogeneous chemical
composition and are spherical with size $a$.  We consider only the
contribution of element $Z$ to $\langle N \rangle$, remembering that
element $Z$ makes up only part of the total mass of the grains.  We
find
\begin{equation}
\langle N \rangle = \frac{4 \rho_Z a}{3 \mu_Z}~,
\end{equation}
where $\rho_Z$ is the mass density of element $Z$ in interstellar
grains and $\mu_Z$ is the atomic mass.  The mass density $\rho_Z$ is
related to the total grain mass density $\rho$ by
\begin{equation}
\rho_Z = \rho \times A_Z \beta_Z \mu_Z \Big/ \sum_{Z} A_Z \beta_Z
\mu_Z~,
\end{equation}
where $\beta_Z$ is the depletion fraction of element $Z$.  The
calculation of $\rho_Z$ takes into account the contribution of all
other elements to the total dust content of the ISM.

For a single element, the number of grains per atom $\xi_{\rm g}$ is
given by
\begin{equation}
\xi_{\rm g} = A_Z \beta_Z \mu_Z \bigg/ \int \frac{dn_{\rm gr}(a)}{da}
\times \rho_Z \times \frac{4}{3} \pi a^3 \times da~.
\end{equation}

Substituting the above relationships into Equation~\ref{eqn:1}, we
find
\begin{equation}
\tau_{\rm grains} = N_{Z} \beta_Z \mu_Z \int^{\infty}_{0}
\frac{dn_{\rm gr}(a)}{da} \pi a^2 [1 - \exp(- \sigma_{\rm gas} \frac{4
\rho_Z a}{3 \mu_Z})] da \bigg/ \int \frac{dn_{\rm gr}(a)}{da} \rho_Z
\frac{4}{3} \pi a^3 da~.
\end{equation}

For our analysis, we are interested in the determining the value of
$f$, the factor which relates the dust cross-section to the gas-phase
cross-section, $\sigma_{\rm grain} = f \sigma_{\rm gas}$.

The grain optical depth $\tau_{\rm grain} = \sigma_{\rm grain}
N_{Z{\rm, grain}}$ where $N_{Z{\rm, grain}}$ is the column density of
element $Z$ found in grains.  The grain column density is given by the
total column density times the depletion factor, $\beta_Z$,
i.e. $N_{Z{\rm, grain}} = \beta_Z N_{Z}$.  Therefore, we can
substitute for the grain optical depth, $\tau_{\rm grains} = f
\sigma_{\rm gas} \beta_{Z} N_{Z}$.  Solving for $f$, we find
\begin{equation}
f = \frac{3 \mu_Z}{4 \rho_Z \sigma_{\rm gas}} \int^{\infty}_{0}
\frac{dn_{\rm gr}(a)}{da} a^2 [1 - \exp(- \frac{4 \rho_Z \sigma_{\rm
gas}}{3 \mu_Z} a)] da \bigg/ \int \frac{dn_{\rm gr}(a)}{da} \times a^3
\times da~,
\end{equation}
after some simplification.

We can now use the above equation to calculate $f$ under various
assumptions.  Our initial assumptions use the ISM abundances and
depletion factors given by \citet{wam00}.  We take $\rho =
1$~g~cm$^{-3}$ and $dn_{\rm gr}(a)/da$ from \citet{mrn77}.  We note
that $\sigma_{\rm gas}$ is a function of energy.

For iron, we assume a single value for $\sigma_{\rm gas}$, which is
the maximum value of the cross-section at the $L_3$ edge.  Using the
maximum value of $\sigma_{\rm gas}$ gives the lowest value of $f$.  We
find $f=0.88$ for iron.

Similarly, for oxygen we can calculate the value of $f$.  The dominant
$K$-shell edge for oxygen occurs at the $1s$-$np$ series limit for the
$S=3/2$ final state (see \jsc).  Using the cross-section value at this
limit to find the value of $f$ yields $f=0.90$.

\clearpage

\end{document}